\documentclass[aps,prc,twocolumn,showpacs,floatfix]{revtex4}
\usepackage{graphicx}
\begin{document}
 
\title{Coulomb Dissociation of $^{15}$C and Radiative Neutron Capture on $^{14}$C}
\author{H. Esbensen}
\affiliation{Physics Division, Argonne National Laboratory, Argonne, Illinois 60439, USA} 
\date{\today}
 
\begin{abstract}
The semiclassical, dynamical description of diffraction dissociation 
of weakly bound nuclei is applied to analyze the decay-energy spectra 
of $^{15}$C that have been measured at 68 MeV/nucleon on a Pb target.
The optical potentials that are used to describe the nuclear interaction 
of $^{15}$C with the target nucleus are realistic because the fits to 
the two measured spectra, one with a small and one with a very large 
acceptance angle, are consistent and of similar quality.
The cross section for the radiative neutron capture on $^{14}$C to 
the 1/2$^+$ ground state of $^{15}$C is deduced from the analysis.
When combined with an estimated contribution from the capture to the
5/2$^+$ excited state of $^{15}$C, an excellent agreement with a recent 
direct capture measurement is achieved.
\end{abstract}
 
\maketitle

\section{Introduction}

The decay-energy spectrum for the breakup of $^{15}$C on a Pb target 
was first reported in Ref. \cite{nakac15a}.
The data were recently analyzed by CDCC (continuum discretized 
coupled-channels) calculations \cite{nunes08}, and it was demonstrated 
that the measured spectrum is consistent with measurements of the 
radiative neutron capture rate on $^{14}$C \cite{reifarth}. 
The final results of the Coulomb dissociation experiment have now been
published \cite{nakac15b}. They include the decay-energy spectrum for
all events and a spectrum for events at forward angles.

It is of interest to investigate whether the two spectra obtained in
Ref. \cite{nakac15b} can be reproduced by a single, consistent theoretical 
description, because the forward angle measurement is primarily due to
Coulomb dissociation, whereas events at large scattering angles are 
influenced by nuclear processes. A more general issue is whether the 
cross section for the radiative neutron capture on $^{14}$C that can 
be deduced from the Coulomb dissociation experiment is consistent with 
the direct capture measurement of Ref. \cite{reifarth}.
  
The method that will be used to analyze the decay-energy spectra of 
$^{15}$C is the semiclassical, dynamical description, where the 
relative motion of the projectile and target follows a classical 
Coulomb trajectory. The breakup reaction, 
$^{15}$C$\rightarrow^{14}$C+n, is calculated quantum mechanically 
by solving the time-dependent Schr\" odinger equation for the 
relative motion of the valence nucleon and the core fragment, in the
time-dependent Coulomb and nuclear fields of the target nucleus. 

The semiclassical model used here was first introduced in Ref. \cite{ebb95} 
as a three-dimensional generalization of a previous two-dimensional 
model \cite{bb94}.  It has been applied in studies of the Coulomb 
dissociation of $^8$B \cite{b8a,b8c} and $^{17}$F\cite{f17} nuclei, and it
has provided a qualitative understanding of some of the phenomena that can 
occur.  For example, intermediate energy Coulomb dissociation experiments are 
commonly analyzed in first-order perturbation theory in terms of E1 transitions
using the so-called far-field approximation and straight line trajectories
\cite{WA-NPA}.
These approximations are rather poor for the Coulomb disassociation of 
weakly bound proton+core systems, partly because the far-field approximation 
breaks down and partly because E2 transitions and higher-order processes
cannot be ignored \cite{b8c,f17}. 
As a consequence, a discrepancy of 15\% was observed between the 
measured radiative capture rate of protons on $^7$Be and the rate inferred 
from Coulomb dissociation experiments \cite{jung}. 
The discrepancy was later resolved by CDCC calculations \cite{Ogata06} 
that included the effects discussed above, in addition to the 
nuclear induced breakup of $^8$B. 

It is surprising that the discrepancy between the direct measurement 
of the cross section for the radiative neutron capture on $^{14}$C
\cite{reifarth} and the first-order analysis of the $^{15}$C Coulomb 
dissociation experiment \cite{nakac15b} is also of the order of 15\%. 
Naively, one would expect that the analysis of the measured 
decay-energy spectra in terms of first-order perturbation theory 
\cite{nakac15b} would result in a much better agreement with the
direct capture measurement because the far-field approximation for 
Coulomb excitation is well justified and the E2 strength is very weak 
for a neutron halo nucleus.
Consequently, the dynamic polarization effect, which is caused by an 
interplay of E1 and E2 transitions \cite{volya}, is much weaker for 
a neutron than for a proton halo nucleus.  
It is therefore of interest to analyze the measured decay-energy within 
the semiclassical, dynamical description, in order to see whether 
the discrepancy with the radiative capture measurement can be reduced.

Part of the explanation for the 15\% discrepancy is that the measured 
neutron capture rate, in addition to the direct capture to the $1/2^+$ 
ground state of $^{15}$C, also includes a small branch to the $5/2^+$
excited state, whereas the Coulomb dissociation experiment can only 
provide information about the capture to the ground state.
Unfortunately, the cross sections for the capture to the ground state
and the excited state have not been measured separately. In this work 
the contribution from the $5/2^+$ branch is estimated to be of the order 
of 4\%, so the issue is what causes the remaining 11\% discrepancy.

Details of the structure input and some results of first-order perturbation
theory are presented in the next section. The basic ingredients of 
the semiclassical method are summarized in Sect. III.
The convergence of the calculations and the analysis of the new 
decay-energy spectra is presented in Sect. IV. The results are used in Sect. V 
to infer the cross section for the radiative neutron capture on $^{14}$C. 
The conclusions are presented in Sect. VI.

\section{Structure models of $^{15}$C}

The single-particle structure associated with the valence neutron
in $^{15}$C is simulated by a Woods-Saxon (WS) potential.  
The depth is adjusted for each partial wave 
so that certain properties of $^{15}$C are reproduced. 
The parameters of the model are shown in Table I.  
The radius and diffuseness, $R$=2.946 fm and $a$=0.5 fm, 
were actually determined in a previous work \cite{terry} by simulating the 
mean-field potential of a realistic Hartree-Fock calculation. The depth
of the s-wave potential, $V_s$, was adjusted to reproduce the 
1.218 MeV neutron separation energy of the $1/2^+$ ground state.  

\begin{table}
\caption{The single-particle structure of $^{15}$C is simulated by a 
Woods-Saxon (WS) potential. The radius $R$ and diffuseness $a$ are 
fixed, and the depth is adjusted for each partial wave, $V_s$, $V_p$ 
and $V_d$, so that certain properties of $^{15}$C are reproduced 
(see the text.) Also shown is the spin-orbit strength, $V_{so}$.}
\begin{tabular} {|c|c|c|c|c|c|}
\colrule
$R$ (fm) & $a$ (fm) & $V_s$ (MeV)& $V_p$ (MeV)& $V_d$ (MeV)& $V_{so}$ (MeV) \\ 
\colrule
2.946  & 0.5    & 55.36 & 55.36 & 52.03 & 4.86 \\ 
\colrule
\end{tabular}
\end{table}

\begin{figure}
\includegraphics [width = 8cm]{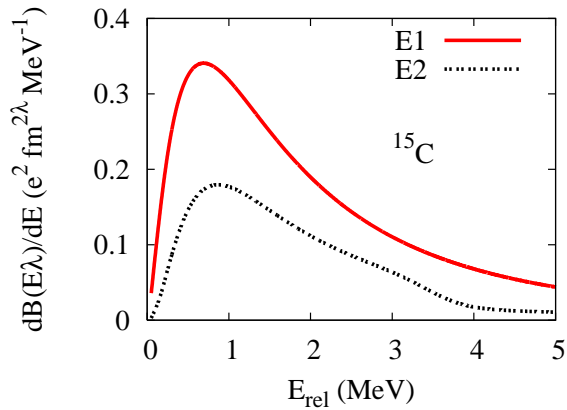}
\caption{(Color online) The calculated dipole and quadrupole responses of 
$^{15}$C are shown as functions of the relative energy $E_{rel}$ of the 
neutron and the $^{14}$C fragment in the final state.}
\label{dbde}
\end{figure} 

The depth of the p-wave potential, $V_p$, is assumed to be 
the same as the depth of the s-wave potential.
This choice was made deliberately in Ref.  \cite{nakac15b}, in order 
not to violate Siegert's theorem which allows one to use charge 
densities instead of current densities when calculating electromagnetic 
matrix elements (see, e.~g., Ref. \cite{greiner}).
The depth of the WS potential for higher, odd-parity partial waves 
is chosen in the following to be the same as for the p-wave potential.

No choice was made in Ref. \cite{nakac15b}  for the d-wave 
potential because it was not needed in the first-order calculations of
dipole excitations performed there. The depth shown in Table I
has been determined in this work so that the 0.478 MeV binding energy 
of the 5/2$^+$ excited state  of $^{15}$C is reproduced. Combined with 
the spin-orbit interaction, this implies that the $3/2^+$ resonance 
is located in close vicinity of the known resonance at 3.56 MeV.

\subsection{First-order Coulomb excitation}

The E1 and E2 strength distributions for $^{15}$C that one obtains are 
illustrated in Fig. \ref{dbde} as functions of the final state relative
energy $E_{rel}$  of the neutron and the $^{14}$C fragment. 
The E2 response has a 
shoulder between 3 and 4 MeV which is a consequence of the 3/2$^+$ 
resonance.  The E2 response does not play any practical role in the
Coulomb dissociation of $^{15}$C because the E2 excitation probability 
is more than three orders of magnitude smaller than the E1 excitation 
probability.  However, the $3/2^+$ resonance does play a role in the 
nuclear induced breakup.

\begin{figure}
\includegraphics [width = 8cm]{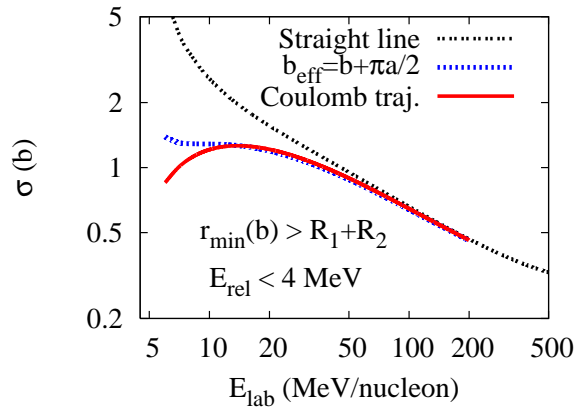}
\caption{(Color online) Coulomb dissociation cross sections for $^{15}$C on a Pb target
as functions of the beam energy. The results of first-order dipole 
excitations are shown for straight-line, corrected straight-line 
($b_{eff}=b+\pi a/2$), and for exact Coulomb trajectories.  
Dipole excitations up to $E_{rel}$ = 4 MeV are included, and the
distance of closest approach is larger than $R_1+R_2$ = 
1.2$(A^{1/3}+A_2^{1/3})$.}
\label{sigcd}
\end{figure} 

Coulomb dissociation experiments at intermediate and high energies
are usually analyzed using the first-order perturbation theory 
that is based on straight-line trajectories \cite{WA-NPA}. 
A simple correction for Coulomb trajectories is often made in this
description. It consists of replacing the minimum impact parameter 
$b_0$ for a given measurement by the effective minimum impact 
parameter $b_0+\pi a/2$, where $a$ is half the collision diameter.
This is actually a very good approximation for first-order dipole 
excitations. This can be seen in Fig. \ref{sigcd} by comparing the
middle dashed curve (labeled $b_{eff}=b+\pi a/2$) to the solid curve, 
which is the first-order Coulomb dissociation cross section obtained 
using exact Coulomb trajectories as described in Ref. \cite{coul}. 

The calculations shown in Fig.  \ref{sigcd} were based on the dipole 
response shown in Fig. \ref{dbde} and included relative energies up 
to 4 MeV and all impact parameters for which the distance of closest 
approach is larger than the sum of the radii, $R_1+R_2$ = 
1.2$(A_1^{1/3}+A_2^{1/3})$.
The top dashed curve is the cross section for straight-line 
trajectories. This is a poor approximation for energies 
below 60 MeV/nucleon. The corrected straight-line trajectory 
approximation (middle curve) is in much better agreement with the 
calculation that is based on exact Coulomb trajectories (the solid curve).
At 68 MeV/nucleon, the two calculations agree within 1\%, whereas the pure
straight-line trajectory calculation (the upper dashed curve) is about 
5\% higher. These results show that the corrected straight-line
trajectory approximation, which was used in the experimental analysis 
\cite{nakac15b}, is well-justified, provided higher-order and nuclear 
processes can be ignored.

\section{The semiclassical method}

The relative motion of projectile and target is assumed to follow a
classical Coulomb trajectory in the semiclassical description of breakup
reactions. The trajectory includes relativistic effects, the most 
important being the kinematics and the determination of the velocity 
from the beam energy. These aspects are discussed in detail in Ref. 
\cite{coul}.

The relative motion of the valence neutron and the $^{14}$C core
is calculated by solving the time-dependent Schr\" odinger equation 
\cite{ebb95}.  The two-body wave function $\Psi({\bf r},t)$, which 
has the $1/2^+$ ground state as initial condition, is expanded 
on the angular momentum eigen states $|ljm\rangle$,
\begin{equation}
\Psi({\bf r},t) = \frac{1}{r} \sum_{ljm}^{\rm L_{max}} \ u_{ljm}(r) \ |ljm\rangle.
\end{equation}
The upper limit L$_{\rm max}$ is the maximum orbital angular momentum
of the expansion.
The radial wave functions $u_{ljm}(r)$ are calculated on a radial grid 
out to 100 fm, with a grid size of 0.1 fm. 
The total wave function is evolved in time by the propagator
\begin{equation}
\Psi(t+\delta t) = 
\bigl[1 - \frac{\delta t}{2i\hbar}H_0\bigr]^{-1} \
\bigl[1 + \frac{\delta t}{2i\hbar}H_0\bigr] \
\bigl[1 + \frac{\delta t}{i\hbar} V_{ext}(t) \bigr] \ \Psi(t),
\end{equation}
where $H_0$ is the two-body Hamiltonian for $^{15}$C, and $V_{ext}(t)$
is the time-dependent interaction of $^{15}$C  with the target nucleus. 
The form of Eq. (2) was originally used in Ref. \cite{kido}. 
The propagator that was used in the earlier work \cite{ebb95} was a 
simplified version of Eq. (2). 

Inserting the expansion (1) into Eq. (2) one obtains the following expression 
for the propagation of the radial wave functions,
$$
u_{ljm}(r,t+\delta t) = 
\bigl[1 - \frac{\delta t}{2i\hbar}h_{lj}\bigr]^{-1} \
\bigl[1 + \frac{\delta t}{2i\hbar}h_{lj}\bigr] \times
$$
\begin{equation}
\Bigl(u_{ljm}(r,t) + \frac{\delta t}{i\hbar} 
\sum_{l'j'm'} \langle ljm|V_{ext}|l'j'm'\rangle \
u_{l'j'm'}(r,t)\Bigr),
\label{propa}
\end{equation}
where $h_{lj}$ is the radial single-particle Hamiltonian
which depends on $(lj)$ through the spin-orbit interaction.
The coupling between the radial wave functions occurs only
through the interaction with the target, whereas the unitary,
intrinsic propagation is diagonal in $|ljm\rangle$. 
The inverse operator, $[1-\delta t/(2i\hbar)h_{lj}]^{-1}$, is 
calculated using the technique described in Appendix B of Ref. \cite{BKN}.

The spin-orbit interaction is included explicitly in the following. 
That makes it possible to treat certain aspects of $^{15}$C in a realistic 
way, such as the 5/2$^+$ bound excited state and the 3/2$^+$ resonance. 
The interaction $V_{ext}$ between the projectile and target consists of 
the core-target Coulomb interaction, the core-target nuclear interaction, 
which is based on the $^{17}$O+Pb optical potential of Ref. \cite{o17pb},
and the Perey-Perey neutron-target interaction \cite{perey}. 
These interactions were chosen here because they were applied in the 
CDCC calculations of Ref. \cite{nunes08}. They were expanded on 
Legendre polynomials,
\begin{equation}
V_{ext}({\bf r},t) = \sum_{\lambda=0}^{\lambda_{max}} 
V_{ext,\lambda}(r,t) \ 
P_{\lambda}(\cos(\theta')),
\label{legendre} 
\end{equation}
where $\theta'$ is the angle between the position ${\bf r}$ of 
the neutron with respect to the $^{14}$C core and the trajectory 
${\bf R}(t)$ of the target with respect to the $^{15}$C projectile.
The maximum value of $\lambda$ in the expansion (\ref{legendre}),
$\lambda_{max}$, was set equal to 2L$_{\rm max}$ so that all of the
necessary multipole components were included in the calculation
of the matrix elements of $V_{ext}$ that appear in Eq. (\ref{propa}).

The semiclassical method is applied to the breakup of $^{15}$C on a Pb target 
at a beam energy  of 68 MeV/nucleon.  The time-evolution of the wave function 
for a given impact parameter $b$ starts when the distance between projectile 
and target is 300 fm with the $1/2^+$ ground state wave function 
as the initial condition. 
The propagation is terminated for practical reasons when the projectile and 
target have re-separated by 100 fm. One could follow the evolution further 
but the result would be useless because the wave function would start to be 
reflected from the outer boundary of the radial box and that produces 
oscillations in the calculated decay-energy spectrum.
 
One way to avoid the problems caused by the reflection of the wave function
from the outer boundary is to use a larger radial box.
This method was used to test the convergence of the calculated decay-energy 
spectrum, and it appears to have converged already at a separation of 100 fm. 
Another way of avoiding the reflection from the outer boundary (which will 
not be used here) is to apply an imaginary potential of a special form that 
acts near the outer boundary; see Ueda et al. \cite{ueda} for details.

\section{Applications to $^{15}$C breakup}

In this section the semiclassical method is applied to analyze the 
decay-energy spectra of $^{15}$C that were measured at 68 MeV/nucleon on a Pb
target \cite{nakac15b}. The parameters of the structure model are shown 
in Table I; they are essentially the same as those that were used in the 
previous first-order analysis of the measurement \cite{nakac15b}.  
Before presenting the results of the 
new analysis of the data, it is useful first to show the dissociation
probabilities one obtains, and also to study the convergence of the 
calculations with respect to the maximum angular momentum L$_{\rm max}$
that is used in the expansion (1) of the $^{15}$C two-body wave function.

\begin{figure}
\includegraphics [width = 8cm]{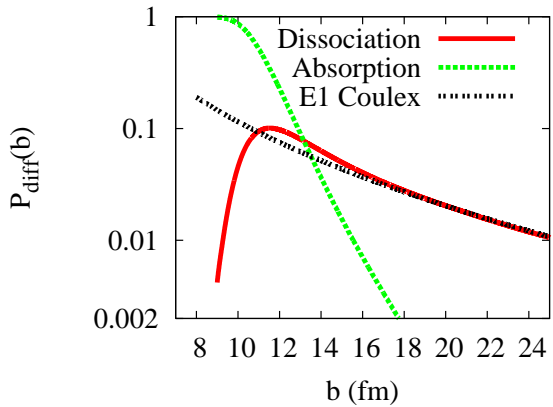}
\caption{(Color online) The diffraction dissociation probability 
$P_{\rm diff}(b)$ for $^{15}$C on a Pb target at 68 MeV/nucleon 
(solid red curve) is shown as a function of the impact parameter $b$ 
and is  compared to the first-order E1 Coulomb dissociation probability 
(black dashed curve).  The steeply falling (green) dashed curve shows
the total absorption probability.}
\label{prob}
\end{figure} 

\subsection{Disassociation probabilities} 

The calculated dissociation probabilities are shown in Fig. \ref{prob}
as functions of the impact parameter. The solid curve is the dynamic 
calculation with L$_{\rm max}$=4. 
The dashed curve (labeled E1 Coulex) is the result 
of the first-order E1 Coulomb excitation, which is here based on Coulomb 
trajectories \cite{coul}. The dynamic calculation is slightly suppressed 
at large impact parameters compared to the first-order calculation.  
In the 11-16 fm impact parameter range, the nuclear induced breakup takes 
over and gives an enhancement compared to the first-order calculation. 
Finally, at impact parameters smaller than 10 fm, the imaginary core-target 
interaction becomes strong and absorbs most of the wave function.  
The steeply falling dashed curve is the total absorption probability, 
which is due to the combined effect of the imaginary part of the 
neutron-target and core-target interactions.  

\subsection{Convergence of the dynamic calculations}

\begin{figure}
\includegraphics [width = 8cm]{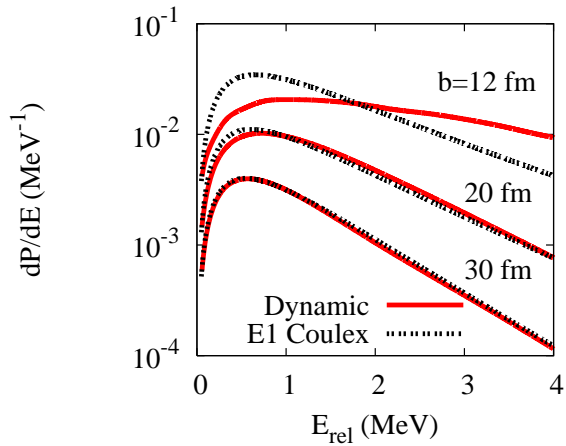}
\caption{(Color online) Decay energy spectra at the impact parameter 
$b$ = 12, 20, and 30 fm.  The solid curves are dynamic calculations, 
whereas the dashed curves are first-order perturbation calculations.}
\label{dpde}
\end{figure}

The decay-energy spectra obtained at three impact parameters are compared 
in Fig. \ref{dpde} to those obtained from first-order E1 transitions. 
The spectra obtained in the dynamic calculations are very broad at 
small impact parameters but they approach the first-order calculation 
at the large impact parameters. 
The decay-energy spectrum, $d\sigma/dE$, will therefore in the following 
be calculated numerically by integrating the dynamic spectra over impact 
parameters less than 30 fm, whereas the contribution from impact parameters 
larger than 30 fm will be estimated by the first-order Coulomb excitation 
spectrum.

\begin{figure}
\includegraphics [width = 8cm]{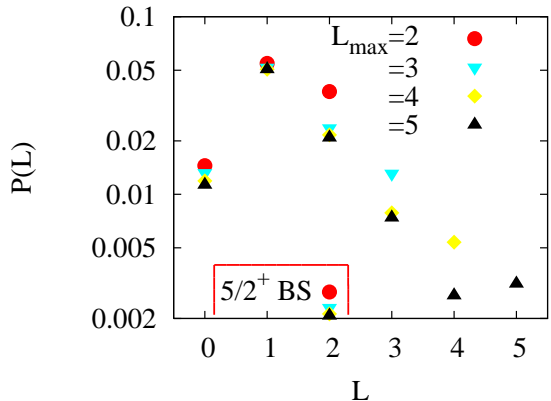}
\caption{(Color online) Occupation probabilities of the continuum as 
function of the orbital angular momentum L of the neutron+$^{14}$C 
system, at the fixed impact parameter $b$=12 fm. 
The probabilities are shown for different values of the maximum 
orbital angular momentum, L$_{\rm max}$. 
The occupation probability of the excited 5/2$^+$ bound state 
(BS) is indicated for L=2 inside the lower left box.}
\label{occ}
\end{figure}

The dependence of the dissociation probability on the maximum angular
momentum L$_{\rm max}$ is illustrated in Fig. \ref{occ} in terms
the continuum occupation probabilities P(L) (summed over $j=L\pm1/2$) 
for a fixed impact parameter of 12 fm.
It is seen that the occupation probability for a fixed value of L 
is largest when L$_{\rm max}$=L, but the value drops and converges 
rather quickly with increasing values of L$_{\rm max}$.
The total dissociation probability is 0.107, 0.102, 0.097 and 0.096
for L$_{\rm max}$ = 2, 3, 4, 5, i.~e., the total probability is
reduced by about 10\% by increasing L$_{\rm max}$ from 2 to 4. 
The occupation probability of the 5/2$^+$ bound state is also shown 
in Fig. \ref{occ} at L=2; it is relatively small and converges to
a value of about 0.002. 

\begin{figure}
\includegraphics [width = 8cm]{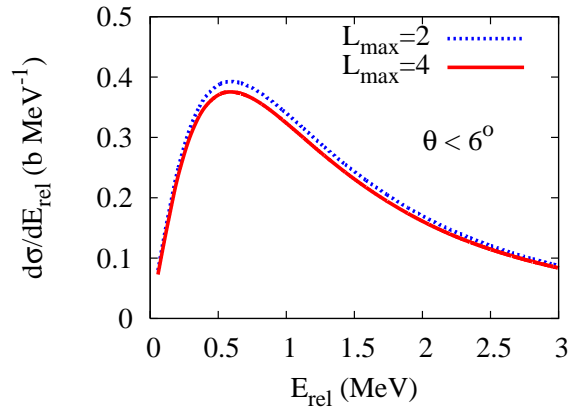}
\caption{(Color online) Decay energy spectra obtained in dynamic calculations
with L$_{max}$ = 2 and 4, respectively. The nominal acceptance angle 
is $6^\circ$ but the calculated spectra have already converged at a 
smaller angle (see text.)}
\label{dsdelm}
\end{figure}

The decay-energy spectra one obtains in the dynamic calculations are 
illustrated in Fig. \ref{dsdelm} for two values of the maximum 
angular momentum, L$_{\rm max}$ = 2 and 4, respectively. 
The peak height decreases from 393 to 375 mb/MeV as L$_{\rm max}$ 
increases from 2 to 4; that is a reduction of 5\%. The peak height is 
not expected to change much by increasing the value of L$_{\rm max}$
further because the reduction is only 1\% when L$_{\rm max}$ is increased 
from 3 to 4.

The nominal maximum center-of-mass acceptance angle in Fig. \ref{dsdelm}
was set to 6$^\circ$ because that is the largest acceptance angle that was
used in the experiment \cite{nakac15b}. The dynamic calculations have actually 
already converged at a smaller scattering angle because of the strong 
absorption at small impact parameters. For example, at the impact parameter 
$b$ = 9 fm, where the dynamic dissociation probability is already quite small, 
the Coulomb scattering angle is $\theta_{\rm cm}$ = 4.9 $^\circ$. 

\subsection{Analysis of the measured spectra}

\begin{figure}
\includegraphics [width = 8cm]{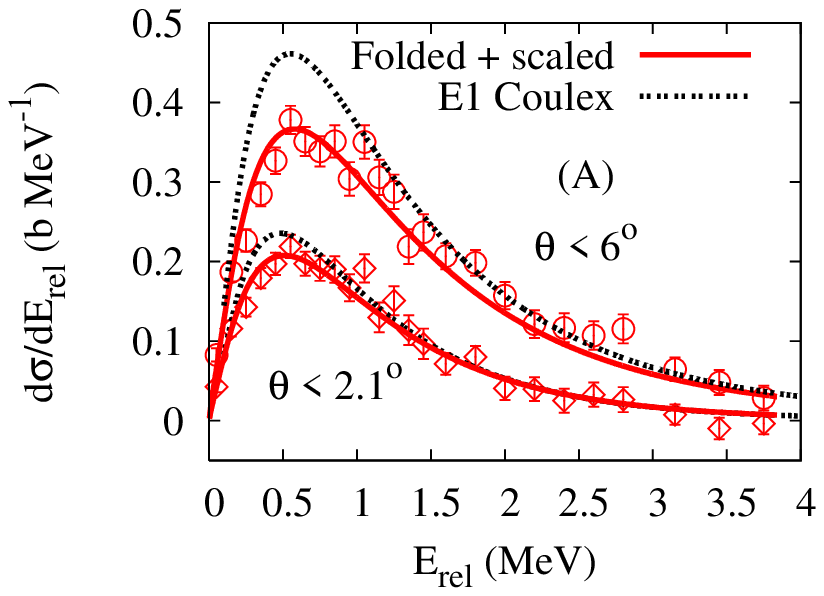}
\includegraphics [width = 8cm]{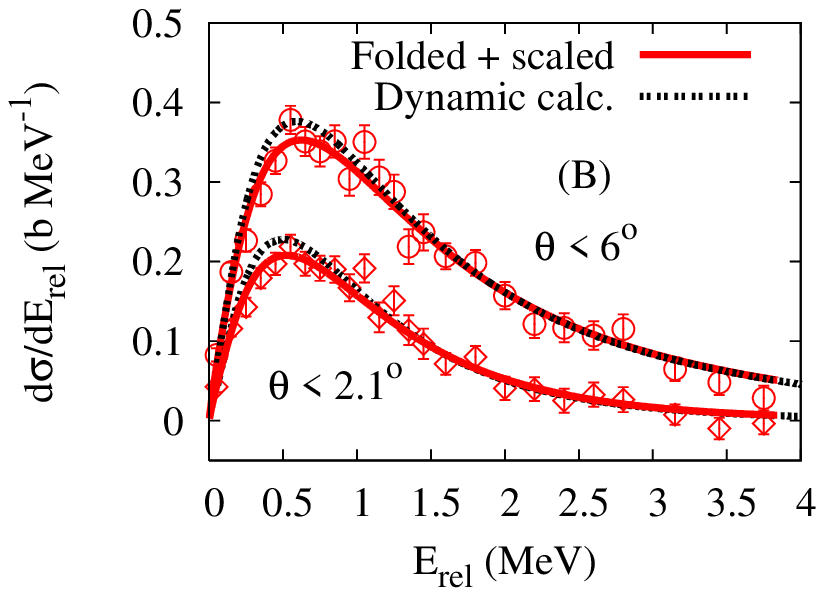}
\caption{(Color online) Decay energy spectra obtained in first-order 
perturbation theory (A) and in the  dynamic calculations with 
L$_{max}$=4 (B) are compared to the data of Ref. \cite{nakac15b}. 
The dashed curves show the calculated spectra. The solid curves 
include the experimental energy 
resolution and have been scaled to give an optimum fit to the data.}
\label{dsdexp}
\end{figure}

The decay-energy spectrum of $^{15}$C at 68 MeV/nucleon on a Pb
target was measured for two center-of-mass acceptance angles,
namely, at 2.1$^\circ$ and 6$^\circ$ \cite{nakac15b}.
The data are compared to the first-order and the 
dynamic calculations in Fig. \ref{dsdexp}A and \ref{dsdexp}B,
respectively. The actual calculations are shown by dashed curves.
The solid curves were obtained by a folding and scaling procedure:
the calculated spectra were first folded with the experimental 
energy resolution, which is a Gaussian with a (1$\sigma$) width
of $\Delta E_{rel}$ = 0.23$\sqrt{E_{rel}}$ \cite{nakac15b}.
The folded spectra were next scaled by the factor $S_{c}$ which
optimizes the $\chi^2$ fit to the data. 
The values of the scaling factors and the associated best values 
of the $\chi^2$ per point are listed in Table II for different 
values of L$_{\rm max}$.

The decay-energy spectra for $\theta<2.1^\circ$ were calculated with 
a sharp cutoff at the impact parameter where the Coulomb scattering 
angle is $\theta$ = 2.1$^\circ$. 
An acceptance angle of 6$^\circ$ does not impose any strict cutoff on 
the dynamic calculation because it converges already at a smaller angle.
The first-order Coulomb excitation calculation, on the other hand, does 
not have such a natural cutoff, except from Coulomb scattering 
but that leads to a very large cross section.
The first-order decay-energy spectrum (E1 Coulex) shown in 
Fig. \ref{dsdexp}A for $\theta<6^\circ$ was therefore determined 
by integrating over all impact parameters for which the minimum 
distance of closest approach is larger than 1.2$(A^{1/3}+A_2^{1/3})$. 
The associated minimum impact parameter for a Coulomb trajectory is 9.7 fm, 
which is a fairly reasonable choice because it falls in the region where 
the absorption in the dynamic calculation sets in (see Fig. \ref{prob}.)
 
\begin{table}
\caption{Analysis of the measured decay-energy spectra 
of $^{15}$C on a Pb target \cite{nakac15b}.
The scaling factor $S_{c}$ and the associated $\chi^2/N$ for the best 
fit to the data up to  $E_{rel}<$ 4 MeV are shown as functions of the 
maximum angular momentum L$_{max}$, and for the two acceptance angles
of the experiment.  Also shown are the cross sections
before and (in parenthesis) after the folding and scaling procedure. } 
\begin{tabular} {|c|c|c|c|c|}
\colrule
          & $\theta<2.1^\circ$ & $\theta<2.1^\circ$ &
            $\theta<6.0^\circ$ & $\theta<6.0^\circ$ \\
L$_{max}$ & $\sigma$ (mb) & $S_{c}$ - $\chi^2/N$
& $\sigma$ (mb) & $S_{c}$ - $\chi^2/N$ \\ 
\colrule
E1 Coulex & 326 [303] & 0.941 - 0.87 & 767 [638] & 0.841 - 2.24 \\ 
 2     & 316 [303] & 0.972 - 0.76 & 750 [696] & 0.939 - 1.15 \\ 
 3     & 316 [303] & 0.972 - 0.77 & 732 [699] & 0.969 - 1.20 \\ 
 4     & 316 [304] & 0.973 - 0.77 & 716 [696] & 0.984 - 1.19 \\  
\colrule
\end{tabular}
\end{table}

The dynamic calculations converge quickly for the smaller acceptance angle.  
This can be seen in Table II where the scaling factor $S_{c}$ that gives
the best fit to the data is independent of L$_{\rm max}$.
At the 6$^\circ$ acceptance angle, the scaling factor increases by almost 
5\% to the value $S_{c}\approx$ 0.98 for L$_{\rm max}$=4; it is not 
expected to increase much further for larger values of L$_{\rm max}$. 
The scaling factors obtained in the two analyses are therefore approximately 
identical and the $\chi^2/N$ is also very reasonable for both acceptance 
angles. 
This implies that the adopted nuclear interactions with the target must be 
realistic because the calculations at large scattering angles are strongly 
influenced by the nuclear interactions, whereas the dissociation at the
smaller acceptance angle is dominated by Coulomb dissociation.

The results of the first-order analysis are shown in the first line 
of Table II. The analysis of the large acceptance angle measurement
is not so interesting because the fit is poor and the necessary 
scaling factor is small and uncertain. The uncertainty stems from 
the crude estimate of the minimum impact parameter.
Although one could possibly choose a better value for the minimum
impact parameter, the fit to the data would still be poor because 
the first-order decay-energy spectra are narrow compared to the 
results of the dynamic calculations at small impact parameters. 
This can be seen in Fig. \ref{dpde}.  

The result of first-order perturbation theory at the smaller acceptance 
angle is much more interesting. Here the $\chi^2/N$ is good and the 
necessary scaling factor does not differ dramatically from the dynamic 
calculations. However, we shall see in the next section that the 4\% 
larger scaling factor of the dynamic calculation, combined with other
corrections, is essential for reaching a good agreement with the 
neutron capture data.

It should be emphasized that the scaling factors $S_{c}$ listed 
in Table II should not be confused with spectroscopic factors. 
The scaling factors were used here as a convenient way to analyze the
data and show how the calculations converge with increasing values of 
L$_{\rm max}$. 

\section{Comparison to radiative capture}

As mentioned in the introduction, the discrepancy between the direct
and indirect measurements of the radiative neutron capture rate on
$^{14}$C is about 15\%. This statement is based on the values of the 
Maxwellian Average Capture (MAC) cross sections, $\sigma_{\rm MAC}$,
that were obtained at the temperature  $kT$ = 23.3 keV.
The definition of the MAC cross section is quoted in Eq. (A1) of the
appendix.
The direct measurement \cite{reifarth} gave the value $\sigma_{\rm MAC}$ 
= 7.1 $\pm$ 0.5 $\mu$b, whereas the indirect, first-order Coulomb 
dissociation method gave the cross section 6.1 $\pm$ 0.5 $\mu$b 
\cite{nakac15b}. One reason for the discrepancy is that the
direct measurement includes a small contribution from the 
capture to the $5/2^+$ excited of $^{15}$C, whereas the indirect 
measurement can only provide information about the capture to the 
$1/2^+$ ground state of $^{15}$C. 

\begin{table}
\caption{MAC cross sections ($kT$ = 23.3 keV) for the radiative 
neutron capture on $^{14}$C to the 1/2$^+$ ground state of $^{15}$C,
to the 5/2$^+$ excited state, and the sum. 
The first line is the prediction of the structure model (with 
$S_{c}$ = 1.) The second and third lines are the results of the 
first-order E1 and the dynamic calculation analysis of the decay-energy 
spectrum with a 2.1$^\circ$ acceptance angle.
The last line is the measured cross section \cite{reifarth}.}
\begin{tabular} {|c|c|c|c|c|c|c|}
\colrule
 Method & $S_{c}$ & $\sigma_{\rm MAC}(1/2^+)$ & 
$\sigma_{\rm MAC}$(5/2$^+$) &
$\sigma_{\rm MAC}$(total) \\
        &          & ($\mu$b) & ($\mu$b) & ($\mu$b) \\ 
\colrule
 Model                &  1     & 6.85   & 0.26 & 7.11   \\
 E1 Coulex           &  0.941 & 6.45(50) & 0.24 & 6.69(50) \\
 Dynamic              &  0.973 & 6.67(50) & 0.25 & 6.92(50) \\
 Exp. \cite{reifarth} &   -    &   -    &   -  & 7.1(5) \\
\colrule
\end{tabular}
\end{table}

In order to investigate in more detail what causes the discrepancy 
between the Coulomb dissociation and radiative capture measurements, 
it is useful to have an estimate of the cross section for the radiative 
capture to the 5/2$^+$ excited state of $^{15}$C.
The prediction of the structure model that was used in the
previous sections is shown in the first line of Table III. 
The total MAC cross section predicted by the model is (accidentally) 
in perfect agreement with the measured 7.1(5) $\mu$b cross 
section \cite{reifarth}.
The contribution from the capture to the 5/2$^+$ excited state 
is almost 4\%, so the discrepancy between the first-order Coulomb 
dissociation analysis of Ref. \cite{nakac15b} and the radiative 
capture measurement is reduced to 11\%.

The second and third lines of Table III show the cross sections 
extracted from the analysis of the decay-energy spectrum that was
measured with the 2.1$^\circ$ acceptance angle (see Table II.)
The cross sections were obtained by multiplying the predictions
of the structure model (the first line of Table III) with the scaling 
factors that gave the best fit to the measured spectrum.
The cross sections have been assigned an uncertainty of 0.5 $\mu$b, 
which roughly reflects the 7\% uncertainty of the Coulomb dissociation 
experiment \cite{nakac15b}. 
The total cross section obtained from the first-order analysis
(second line) is smaller than but not inconsistent with the 
capture measurement.  
The cross section extracted from the dynamic calculation (third line) 
is in excellent agreement with the measurement. 
Thus it is the combined effect of dynamic processes and the estimated 
contribution from the capture to the 5/2$^+$ state that makes it 
possible to achieve consistency between the Coulomb dissociation 
and the direct capture measurements.

The MAC cross section for the radiative capture to the 1/2$^+$ ground 
state of $^{15}$C that was obtained in Ref. \cite{nakac15b} is 
6.1(5) $\mu$b.  That is about 5\% smaller than the 6.45(50) $\mu$b 
cross section shown in the second line of Table III. 
The two results should in principle be identical because they are based
on the same data set (the decay-energy spectrum at 2.1$^\circ$) and they
were both obtained using first-order perturbation theory.
In this connection, it is very interesting that the two analyses appear
to give the same dipole strength distribution. For example, the peak 
height of the extracted distribution shown by the dashed curve in Fig. 2 
of Ref. \cite{nakac15b} is 0.32 e$^2$fm$^2$/MeV. 
The same peak height is obtained from the dipole response shown Fig. 1 
when multiplied with the (E1 Coulex) scaling factor of Table II:
0.34 $\times$ 0.941 = 0.32 e$^2$fm$^2$/MeV.
The 5\% discrepancy mentioned above must therefore have developed in the
calculation of the MAC cross section from the extracted dipole strength 
distribution (see appendix.)


\begin{figure}
\includegraphics [width = 8cm]{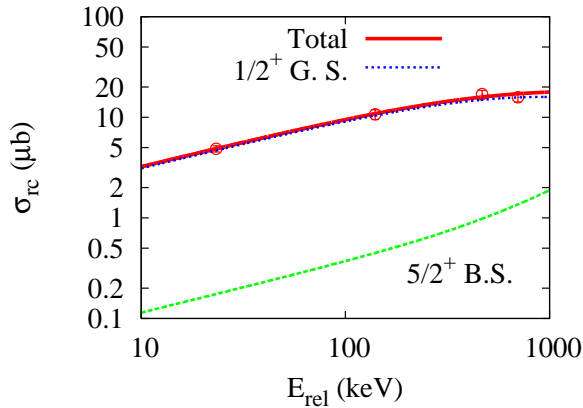}
\caption{(Color online)
Calculated cross sections for the radiative neutron capture on 
$^{14}$C to the 1/2$^+$ ground state (G.S.), the 5/2$^+$ bound 
state (B.S.) of $^{15}$C, and the sum (Total) are shown as 
functions of the $^{14}$C+neutron relative energy, $E_{rel}$.
The measured cross sections are from Ref. \cite{reifarth}.}
\label{radc}
\end{figure} 

The radiative capture cross sections predicted by the 
structure model defined by Table I are shown in Fig. \ref{radc} 
as functions of the neutron center-of-mass energy. 
The figure shows the contributions from the capture to the 1/2$^+$ 
ground state and to the 5/2$^+$ excited state of $^{15}$C.
The solid curve is the sum (total) which can be compared to the 
cross sections that were measured in Ref. \cite{reifarth}. 
The model is in very good agreement with the data.
The best prediction obtained from the analysis of the Coulomb
dissociation experiment is a factor of 0.973 times the model 
prediction which is also in good agreement with the data.

The experimental cross section shown at the lowest energy 
(23.3 keV) in Fig. \ref{radc} is 4.86(34) $\mu$b. 
This value was determined from the published 7.1(5) $\mu$b MAC cross 
and the simple relation, Eq. (A3), derived in the appendix. 
A different value, namely 5.2(3) $\mu$b, was quoted in Ref. 
\cite{reifarth} but that value has been discarded here because it does 
not agree with the prediction of Eq. (A3), which was derived from the 
structure model used here and eighth other structure models that were 
considered in Ref. \cite{timo}.

\section{Conclusions}

The semiclassical, dynamical description of the dissociation of
weakly bound nuclei was applied to analyze the decay-energy spectra 
of $^{15}$C that have been measured in reactions with a Pb target.
The two-body structure model of $^{15}$C, which was partly 
developed previously, was supplemented with a model for d-waves, so 
that the energies of the $5/2^+$ bound state and the $3/2^+$ 
resonance could be simulated. Standard neutron-target and core-target 
optical potentials were employed to describe the nuclear interaction 
of $^{15}$C with the target nucleus.

The two spectra that were  measured were obtained with a small 
and a large acceptance angle, respectively. The calculated spectrum 
for the small acceptance angle is dominated by Coulomb dissociation,
whereas the spectrum calculated with the large acceptance angle
is strongly influenced by the nuclear interaction.
The effect of the nuclear interaction is to produce broad decay-energy 
spectra at large scattering angles, and it also determines 
the magnitude of the spectrum for the large acceptance angle because 
of a strong absorption it produces at small impact parameters.

The analysis of the measured decay-energy spectra shows that the 
optical potentials that were used must be fairly realistic because 
the scaling factors that give the best fit to the spectra at the small 
and large acceptance angles are essentially the same.
The analysis also shows that the adopted structure model must be
fairly realistic because the best fit to the data is achieved with
a scaling factor that is very close to one.

Having achieved a comprehensive description of the measured decay-energy 
spectra, the structure model was applied to calculate the
cross sections for the radiative neutron capture on $^{14}$C, not 
only to the ground state but also to an excited state of $^{15}$C.
The total capture cross section predicted by the model is (accidentally) 
in perfect agreement with the recently measured, low-energy capture
cross section, and it also accounts for the neutron capture data at 
higher energies. That implies that the measured decay-energy spectra 
and the measured neutron capture cross sections are consistent within 
a few percent, when the decay-energy spectra are analyzed in
semiclassical, dynamical description. 

Good agreement between the dissociation and neutron capture measurements 
was also recently achieved within the CDCC description \cite{nunes08}.
The analysis of the dissociation data was based on structure models
that are slightly different from the model used here, in 
particular with respect to d-waves. A comparison of
the semiclassical and CDCC calculations will be made in the near 
future \cite{capel}.

{\bf Acknowledgments} 
The author is grateful to F. Nunes for discussions and to T. Nakamura
for providing the data and information about the experiment.
This work was supported by the U.S. Department of Energy,
Office of Nuclear Physics, under Contract No. DE-AC02-06CH11357.

\section{Appendix: MAC cross section}

The Maxwellian Averaged Capture (MAC) cross section at the temperature 
$kT$ is defined by (see Eq. (5) of Ref. \cite{beer})
$$
\sigma_{\rm MAC} = \frac{2}{\sqrt{\pi}}
\int \frac{dE}{kT} \ \sigma_{rc}(E) \ 
\frac{E}{kT} \ exp(-\frac{E}{kT}), \eqno(A1)
$$
in terms of the radiative capture cross section $\sigma_{rc}(E)$,
which can be derived from the dipole strength distribution. 
For example, the cross section for the E1 radioactive neutron capture 
to the  1/2$^{+}$ ground state of $^{15}$C is 
$$\sigma_{rc}(E) = \frac{4(2\pi)^3}{9(\hbar c)^3} \
\frac{\hbar^2}{2\mu_n} \  
\frac{(S_n+E)^3}{E} \ 
\frac{dB(E1)}{dE}, \eqno(A2)
$$
where 
$dB(E1)/dE$ is the dipole excitation strength, $S_n$ is the neutron 
separation energy, $\mu_n$ the neutron-$^{14}$C reduced mass,
and $E$ is the relative energy of the neutron and 
the $^{14}$C fragment. A more general expression can be found, for
example, in section 3.2 of Ref. \cite{b8a}.

Inserting into Eq. (A1) the cross section for the capture to the 1/2$^+$ 
ground state of $^{15}$C calculated in the model defined by the parameters 
in Table I, one obtains the following relation between the MAC cross section
at the temperature $kT$ = 23.3 keV 
and the capture cross section at the energy $kT$ = 23.3 keV,
$$
\sigma_{\rm MAC} \approx 1.46 \ \sigma_{rc}(kT).
\eqno(A3)
$$
(If the small capture cross section to the 5/2$^+$ bound state is also 
included, one obtain essentially the same relation, namely, with the 
factor 1.460 replaced by 1.462.)

Daniel Baye derived the following expression \cite{timo}
$$
\sigma_{rc}(E) \approx C_0 \sqrt{E}
(1 + s_1 E + s_2 E^2),
\eqno(A4)
$$
which gives a very good parametrization of the low-energy cross section
for the radiative neutron capture from a p-wave to an s-wave bound state.
Inserting this expression into Eq. (A1) one obtains the expression
$$
\sigma_{\rm MAC} = \frac{3}{2} \ \sigma_{rc}(kT) \
\frac{1+\frac{5}{2} s_1 kT + \frac{35}{4} s_2 (kT)^2}
{1 + s_1 kT + s_2 (kT)^2}.
\eqno(A5)
$$
Eight different structure models of $^{15}$C were considered in
Ref. \cite{timo} and they all give the same relation, Eq. (A3), between
the capture cross section at 23.3 keV and the MAC cross section at the
temperature $kT$ = 23.3 keV, when the parameters $s_1$ and $s_2$ of
the models are inserted into Eq. (A5).

The cross section for the radiative capture to the 1/2$^+$ ground state
of $^{15}$C predicted by structure model defined in Table I can be
accurately parametrized by Eq. (A4) (at least up to 1 MeV) with 
$C_0\sqrt{kT}$ = 4.77 $\mu$b, 
$s_1$ = --0.783 MeV$^{-1}$ and $s_2$ = 0.298 MeV$^{-2}$.


\begin{thebibliography}{99}
\bibitem{nakac15a} T. Nakamura {\it et al.}, Nucl. Phys. A {\bf 722}, 301c (2003).
\bibitem{nunes08} N. C. Summers and F. M. Nunes Phys. Rev. C {\bf 78}, 011601(R) (2008);
ibid. {\bf 78} 069908(E) (2008).
\bibitem{reifarth} R. Reifarth {\it et al}., Phys. Rev. C {\bf 77}, 015804 (2008).
\bibitem{nakac15b} T. Nakamura {\it et al}., Phys. Rev. C {\bf 79}, 035805 (2009).
\bibitem{ebb95} H. Esbensen, G. F. Bertsch, and C. A. Bertulani,
                Nucl. Phys. {\bf A581}, 107 (1995).
\bibitem{bb94} C. A. Bertulani and G. F. Bertsch,
                Phys. Rev. C {\bf 49}, 2839 (1994).
\bibitem{b8a} H. Esbensen and G. F. Bertsch, 
              Nucl. Phys. A {\bf 600}, 37 (1996);
              Phys. Rev. C {\bf 66}, 044609 (2002).
\bibitem{b8c} H. Esbensen, G. F. Bertsch, and K. A. Snover,
              Phys. Rev. Lett. {\bf 94}, 042502 (2005).
\bibitem{f17} H. Esbensen and G. F. Bertsch, Nucl. Phys. A {\bf 706}, 383 (2002).
\bibitem{WA-NPA} A. Winther and K. Alder, Nucl. Phys. {\bf A319}, 518 (1979).
\bibitem{jung} A. R. Junghans {\it et al}., Phys. Rev. C {\bf 68}, 065803 (2003).
\bibitem{Ogata06} K. Ogata, S. Hashimoto, Y. Iseri, M. Kamimura, and M. Yahiro,
                  Phys. Rev. C {\bf 73}, 024605 (2006).
\bibitem{volya} Alexander Volya and Henning Esbensen, 
                Phys. Rev. C {\bf 66}, 044604 (2002).
\bibitem{terry} J. R. Terry, D. Bazin, B. A. Brown, J. Enders, T. Glasmacher,
                P. G. Hansen, B. M. Sherrill, and J. A. Tostevin,
                Phys. Rev. C {\bf 69}, 054306 (2004).
\bibitem{greiner} J. M. Eisenberg and W. Greiner, 
        {\it Excitation Mechanisms of the Nucleus, Nuclear Theory 2}
        (3rd revised edition, North Holland, Amsterdam (1988)). 
\bibitem{coul} H. Esbensen, Phys. Rev. C {\bf 78}, 024608 (2008).
\bibitem{kido} T. Kido, K. Yabana, Y. Suzuki, Phys. Rev. C {\bf 50}, R1276 (1994);
               Phys. Rev. C {\bf 53}, 2296 (1996).
\bibitem{BKN} P. Bonche, S. Koonin, J. W. Negele, 
              Phys. Rev. C {\bf 13}, 1226 (1976).
\bibitem{o17pb} N. Fukuda {\it et al}., Phys. Rev. C {\bf 70}, 054606 (2004).
\bibitem{perey} C. M. Perey and F. G. Perey. At. Nucl. Data Tables, {\bf 17}, 1 (1976).
\bibitem{ueda} M. Ueda, K. Yabana and T. Nakatsukasa, Phys. Rev. C {\bf 67}, 014606 (2003).
\bibitem{timo} N.K.Timofeyuk,
D.Baye, P.Descouvemont, R.Kamouni, 
and  I.J.Thompson, Phys. Rev. Lett. {\bf 96}, 162501 (2006).
\bibitem{capel} P. Capel and F. Nunes (private communications).
\bibitem{beer} H. Beer, F. Voss, and R. R. Winters, Astrophys. Jour. Suppl. 
               {\bf 80}, 403 (1992).
\end{thebibliography}
\end{document}